\documentstyle[11pt]{article}
\normalbaselines
\begin{document}
\bibliographystyle{unsrt}

\newtheorem{theorem}{ Theorem}
\newtheorem{lemma}{Lemma}
\newtheorem{proposition}{Proposition}


 
\def\bea*{\begin{eqnarray*}}
\def\eea*{\end{eqnarray*}}
\def\ba{\begin{array}}
\def\ea{\end{array}}
% --------------------------------------------------------------
% If you want to see the names of the equations and references
% set below \count1=0 otherwise \count1=1
% --------------------------------------------------------------
\count1=1
% --------------------------------------------------------------
\def\be{\ifnum \count1=0 $$ \else \begin{equation}\fi}
\def\ee{\ifnum\count1=0 $$ \else \end{equation}\fi}
\def\ele(#1){\ifnum\count1=0 \eqno({\bf #1}) $$ \else \label{#1}\end{equation}\fi}
\def\req(#1){\ifnum\count1=0 {\bf #1}\else \ref{#1}\fi}
\def\bea(#1){\ifnum \count1=0   $$ \begin{array}{#1}
\else \begin{equation} \begin{array}{#1} \fi}
\def\eea{\ifnum \count1=0 \end{array} $$
\else  \end{array}\end{equation}\fi}
\def\elea(#1){\ifnum \count1=0 \end{array}\label{#1}\eqno({\bf #1}) $$
\else\end{array}\label{#1}\end{equation}\fi}
\def\cit(#1){
\ifnum\count1=0 {\bf #1} \cite{#1} \else 
\cite{#1}\fi}
\def\bibit(#1){\ifnum\count1=0 \bibitem{#1} [#1    ] \else \bibitem{#1}\fi}
\def\ds{\displaystyle}
\def\hb{\hfill\break}
\def\comment#1{\hb {***** {\em #1} *****}\hb }

\newcommand{\TZ}{\hbox{\bf T}}
\newcommand{\MZ}{\hbox{\bf M}}
\newcommand{\ZZ}{\hbox{\bf Z}}
\newcommand{\NZ}{\hbox{\bf N}}
\newcommand{\RZ}{\hbox{\bf R}}
\newcommand{\CZ}{\,\hbox{\bf C}}
\newcommand{\PZ}{\hbox{\bf P}}
\newcommand{\QZ}{\hbox{\bf Q}}
\newcommand{\HZ}{\hbox{\bf H}}
\newcommand{\EZ}{\hbox{\bf E}}
\newcommand{\GZ}{\,\hbox{\bf G}}

\vbox{\vspace{38mm}}
\begin{center}
{\LARGE \bf 
Algebraic Geometry in Discrete  Quantum \\ [2mm]
Integrable Model and Baxter's T-Q Relation  
}\footnote{Presented at the 10th Colloquium
"Quantum groups and integrable systems",
Prague, 21-23 June, 2001. }\\[10 mm] Shao-shiung
Lin
\\ {\it Department of Mathematics , \  
Taiwan University \\ Taipei, Taiwan \\
( email: lin@math.ntu.edu.tw ) }\\[3 mm]
Shi-shyr Roan\footnote{Supported in part by   
NSC grant 
89-2115-M-001-037 of Taiwan.}
\\ {\it Institute of Mathematics , \ 
Academia Sinica \\  Taipei , Taiwan \\
(email: maroan@ccvax.sinica.edu.tw ) } \\[25mm]
\end{center}

\begin{abstract}
We study the diagonalization problem of certain
discrete quantum integrable models 
by the method of Baxter's $T$-$Q$ relation from
the algebraic geometry aspect. Among those the
Hofstadter type model  (with the rational
magnetic flux), discrete quantum pendulum and
discrete sine-Gordon model are our main
concern in this report. By the quantum inverse
scattering method, the Baxter's $T$-$Q$ relation
is formulated on the associated spectral curve, a
high genus Riemann surface in general,
arisen from the study of the spectrum
problem of the system.  In the case of degenerated
spectral curve where the  spectral
variables lie on rational curves, we obtain the
complete and explicit solution of the $T$-$Q$
polynomial equation associated to the model, and
the intimate relation between the Baxter's $T$-$Q$
relation and algebraic Bethe Ansatz  is clearly
revealed. The algebraic geometry of a general
spectral curve attached to the model and certain
qualitative properties of solutions of the 
Baxter's $T$-$Q$  relation  are  discussed
incorporating the physical consideration.
\end{abstract}

$$
\begin{array}{ll}
PACS:& 02.10. {\rm Rn}, \ 03.65. {\rm Fd}, \ 5.30,
\ 75.10{\rm J}.
\\ Key \ words: & {\rm  Baxter
} \ T-Q \ {\rm  relation, \ Hofstadter \ model, \
Discrete
\ quantum } \\
& {\rm pendulum, \ Discrete \ sine-Gordon \ model.
}
\end{array}
$$

\section{Introduction}
This report is a review based on our results in
\cite{LR99, LR01}, the relevant work on the
subject can be found in the references
therein, especially on the connection
with \cite{BKP, FK, KKS}.  We
consider the diagonalization problem of certain
discrete quantum integrable chains by the method
of Baxter's $T$-$Q$ relation \cite{B} within the
context of quantum inverse scattering method
\cite{F, KS}. The models we are going to discuss
here are the Hofstadter type model  (with the
rational magnetic flux), discrete quantum
pendulum and discrete (quantum) sine-Gordon
model.  Our approach to the eigenvalue problem of
these Hamiltonians is from the algebraic geometry
aspect by taking the algebraic structure of
rational functions of the underlying spectral
curve into account, then formulating the problem
into the   Baxter's $T$-$Q$ relation \cite{B}; a
method often not only more general, but also
mathematically tractable than the usual Bethe
Ansatz technique. This is indeed the case when
the spectral curve degenerates into rational
curves for these models, where the complete
solutions of Baxter's
$T$-$Q$ polynomial equation are obtained, and
both the qualitative and quantitative features of
eigenvalues and eigenfunctions have been studied
in details incorporating the physical
consideration. The method starts with the transfer
matrix technique by encoding the Hamiltonian into
the transfer matrix of a fixed finite size $L$,
which arises from a Yang-Baxter solution for the
$R$-matrix of XXZ model. The Baxter's $T$-$Q$
relation is constructed from the action of 
transfer matrices on the Baxter vacuum state over
the spectral curve.  The  Hofstadter type
Hamiltonian is on
$L=3$, and  the discrete quantum pendulum and
sine-Gordon model are on
$L=4$, (see (I) (II) (III) of Sect 2). 
The mathematical treatment to solutions of these
Baxter's $T$-$Q$ polynomial equations takes
explicit advantage of special features only
presented in
$L=3,4$. For a general spectral curve, it is a
Riemann surface of a very high genus, however
with a close link with elliptic curves for both
cases. The qualitative analysis in the geometry of
spectral curves has been made in respect to
Baxter's
$T$-$Q$ relation. When the spectral curve is
totally degenerated into rational curves, the
Baxter's $T$-$Q$ relation is descended to a
polynomial equation for an arbitrary  size
$L$. In this situation, though the geometry of
the spectral curve becomes trivial, the
determination of the solutions inevitably
requires the subtle analysis of Baxter vacuum
state to extract the essential data for
polynomials involved, also the necessary
algebraic study of a certain
"over-determined" system of 
$q$-difference  equations for a root of unity
$q^N=1$, which has still been a difficult problem
for an arbitrary finite size $L$. For $L=3,
4$ in accordance with  models concerning us
here, a sound mathematical derivation of 
solutions of the Baxter's $T$-$Q$ polynomial
equation has been carried out in \cite{LR99,
LR01}. The relationship between the Baxter's
$T$-$Q$ polynomial equation and algebraic Bethe
Ansatz becomes clearer in this scheme.

{\bf Notations. }  In this
article, 
$\ZZ, \RZ, \CZ$ will denote 
the ring of integers, real, complex numbers
respectively, $\NZ = \ZZ_{>0}$, $\ZZ_N=
\ZZ/N\ZZ$,  and ${\rm i} = \sqrt{-1}$. 
For the $N$-dimensional  vector space 
$\CZ^N$, we will present a vector 
$v \in \CZ^N$ as a sequence of
coordinates, $(v_k \ | \  k
\in \ZZ)$, with the $N$-periodic condition, $v_k =
v_{k+N}$, i.e., $v = (v_k)_{k
\in \ZZ_N}$.  
The standard basis of $\CZ^N$ will be denoted by 
$|k\rangle$, with the dual basis of $\CZ^{N*}$ by
$\langle k|$ for $k \in \ZZ_N$. 
For a positive integers $n$, we denote
$\stackrel{n}{\otimes} \CZ^N$ the tensor
product of $n$-copies of the vector space $\CZ^N$.
We use the
notation of ${\sf q}$-shifted  factorials,
$$
(a ; {\sf q} )_0 = 1, \ \ \ \
(a ; {\sf q} )_n = 
(1-a)(1-a {\sf q}) \cdots (1-a
{\sf q}^{n-1}) , \ \ \ \ n \in \NZ  \ .
$$

\section{Baxter's T-Q Relation on the Spectral
Curve}
Throughout this report, $N$ will always denote an
odd positive  integer with $M= [\frac{N}{2}]$, 
$$
N= 2M+1 \ , \ \ \ M \geq 1 \ ,
$$ 
and $\omega$ is a primitive
$N$-th root of unity,  $q:=
\omega^{\frac{1}{2}}$ with   
$q^N = 1$, i.e., 
$q = \omega^{M+1 }$.

Let $Z, X$ be a pair of operators generating 
the Weyl algebra and
$Y:=ZX$ with the  Weyl commutation
relation and the
$N$-th power identity, 
$$
ZX= \omega XZ , \ \ \ XY = \omega^{-1} YX , \
\ \ YZ = \omega^{-1} ZY \ ; \ \ \  Z^N = X^N = Y^N
= I
\ .
$$
The canonical  
representation of the Weyl algebra is the
unique irreducible one on $\CZ^N$ with the
expression:
$$
Z(v)_k = \omega^n v_k \ ,
\ \  X(v)_k = v_{k-1} \ , \
\ \ Y(v)_k = \omega^k
v_{k-1} \ \ \ {\rm for} \ \ v=(v_k) \in \CZ^N .
$$
Using  $X, Y, Z$ and the identity
operator,  there is  a solution of the
Yang-Baxter equation for a  slightly modified 
$R$-matrix of the XXZ-model \cite{FK} with
the 
$L$-operator given by the following matrix of the
operator-valued entries acting on the quantum
space $\CZ^N$, depending on the parameter
$h=[a:b:c:d]$ in the projective 3-space
$ \PZ^3$,      
\be
L_h (x) = \left( \begin{array}{cc}
       aY  & xbX  \\
        xcZ &d    
\end{array} \right) \ , \ \ x \in \CZ \ .
\ele(L)
It satisfies the Yang-Baxter relation, 
$$
R(x/x') (L_h (x) \bigotimes_{aux}1) ( 1
\bigotimes_{aux} L_h (x')) = (1
\bigotimes_{aux} L_h (x'))(L_h (x)
\bigotimes_{aux} 1) R(x/x') 
$$
where 
$$
R(x) = \left( \begin{array}{cccc}
        x\omega-x^{-1}  & 0 & 0 & 0 \\
        0 &\omega(x-x^{-1}) & \omega-1 &  0 \\ 
        0 & \omega-1  &x-x^{-1} & 0 \\
     0 & 0 &0 & x\omega-x^{-1}     
\end{array} \right) \ .
$$
The above relation is related to the
Yang-Baxter solution for the ${\sf R}$-matrix of
XXZ-model in
\cite{IK} on the study of  sine-Gordon lattice
model by using 
the Weyl operators $U, V$: $UV = q^{-1} VU$,
$ U^N=V^N =1$, 
$$
L^{\ast}_h (x) = \left( \begin{array}{cc}
       aqU  & xbV^{-1}  \\
        xcV &dU^{-1}    
\end{array} \right) \ , \ \ 
{\sf R}(x) = \left( \begin{array}{cccc}
        x q -x^{-1}q^{-1}  & 0 & 0 & 0 \\
        0 & x-x^{-1} & q-q^{-1} &  0 \\ 
        0 & q-q^{-1}  & x-x^{-1} & 0 \\
     0 & 0 &0 &  x q -x^{-1}q^{-1}    
\end{array} \right) \ .
$$
In fact, with the identification $
Z= VU,  X= V^{-1}U$, or equivalently, $ U =
q^{\frac{-1}{2}}Y^{\frac{1}{2}} ,  V =
q^{\frac{1}{2}} Z Y^{\frac{-1}{2}} $, the
relation between
$L_h(x)$ and 
$L^{\ast}_h(x)$  is given by $
L^{\ast}_h (x) =  L_h (x)Y^{\frac{-1}{2}}
q^{\frac{1}{2}}$.
By the matrix-product on 
auxiliary  spaces and tensor-product of 
quantum spaces, the
$L$-operator for a finite size $L$ with
the period boundary condition and the
parameter
$\vec{h}= (h_0,
\ldots, h_{L-1}) \in  (\PZ^3)^L $, $
L_{\vec{h}}(x) =
\bigotimes_{j=0}^{L-1}L_{h_j}(x) $ 
again satisfies the Yang-Baxter equation, hence it
gives rise to the commuting family of the transfer
matrices 
$$
T_{\vec{h}} (x) = {\rm tr}_{aux} ( L_{\vec{h}}(x))
\ , \ \ \ x \in \CZ \ .
$$
The same conclusion holds for
$L^{\ast}_{\vec{h}} (x)$ and $ 
T^{\ast}_{\vec{h}} (x)$, and the following
relation holds for these two families  of transfer
matrices: 
\be
T^{\ast}_{\vec{h}} (x)   = T_{\vec{h}}
(x)D^{\frac{-1}{2}} \ \ \ \ 
{\rm where } \ \ \ D: = q^{-L}
\stackrel{L}{\otimes} Y .
\ele(TT*)
The transfer matrix $T_{\vec{h}}(x)$ acts on the
quantum space $\stackrel{n}{\otimes} \CZ^N$, also
its dual space
$\stackrel{n}{\otimes} \CZ^{N *}$. As
$T_{\vec{h}}(x)$ for $x \in \CZ$ form a commuting
family of operators, all the
operators $T_{\vec{h}}(x)$ of
$\stackrel{L}{\otimes}\CZ^{N*}$ can be
simultaneously diagonalized by common
eigenvectors
$\langle
\varphi| $ with the
eigenvalue $\Lambda(x) (= \Lambda_\varphi(x)) \in
\CZ[x]$. The quest of spectrum  of
$T_{\vec{h}}(x)$ ( or $T_{\vec{h}}^\ast(x)$) is
the problem for our main concern in this work. 
It is not hard too see that $T_{\vec{h}}(x)$ can
be expressed as an even polynomial of
$x$ with operator-coefficients:
$$
T_{\vec{h}}(x) =
\sum_{j=0}^{[\frac{L}{2}]} T_{2j} x^{2j} \ , \ \
\ \ [T_{2j}, T_{2k} ] = 0 \  \ 
$$
with $T_0  = ( q^L
\prod_{j=0}^{L-1}a_j ) D + 
\prod_{j=0}^{L-1} d_j$. This implies that the
eigenvalue $\Lambda(x)$ is an even polynomial with
degree $\leq [\frac{L}{2}]$ and $\Lambda(0)=
q^s \prod_{j=0}^{L-1}a_j + 
\prod_{j=0}^{L-1} d_j $ for $s \in \ZZ_N$.
For
$L=3, 4$, the expressions of $T_{2j}$ are listed
as follows:
$$
\begin{array}{lll}
 L=3 , & T_2  =& \ c_0b_1d_2 Z\otimes X \otimes I +
d_0c_1b_2 I \otimes  Z \otimes X+ 
 b_0d_1c_2 X \otimes I
\otimes Z + (a_jY \leftrightarrow d_j ,
 b_j X \leftrightarrow  c_j Z ) ; \\
L=4, & T_2 =& d_0d_1c_2b_3 1 \otimes 1
\otimes Z \otimes X  +
b_0d_1d_2c_3 X \otimes 1 
\otimes 1 \otimes Z + c_0b_1 d_2 d_3 Z \otimes X
\otimes 1 \otimes 1  + \\
 & & d_0 c_1 b_2 d_3 1 \otimes Z
\otimes X \otimes 1 + d_0c_1
a_2b_3 1 \otimes Z \otimes Y \otimes X + c_0a_1
b_2d_3 Z \otimes Y \otimes X \otimes 1 + \\
 & &  (a_jY \leftrightarrow d_j ,
 b_j X \leftrightarrow  c_j Z ) \ , \\
& T_4 =& b_0c_1b_2c_3  D C^{-1}
+ c_0b_1 c_2b_3 C \ , \ \ C:= Z \otimes X \otimes
Z \otimes X \ .
\end{array}
$$
A general scheme to study the
diagonalization problem of $T_{\vec{h}}(x)$ by a 
"difference-like" equations over an  algebraic
curve via the Baxter vacuum state  has been
proposed in \cite{FK, LR99}, of which we are
going to describe the method as follows.

In computing the spectra of 
$T_{\vec{h}}(x)$, one can apply the  gauge
transform on 
$L_{h_j}(x) $ in a special form,
$$
\widetilde{L}_{h_j}(x, \xi_j, \xi_{j+1}) = A_j
L_{h_j}(x) A_{j+1}^{-1}
\ ,\ \ A_j = \left( \begin{array}{lc}
1 &  \xi_j-1\\
1 &  \xi_j 
\end{array}\right)  \ \  0 \leq j \leq L-1 ,
$$
with $A_{L}: = A_0$. We have 
$$
\widetilde{L}_{h_j} (x, \xi_j, \xi_{j+1})  = \left(
\begin{array}{ll} F_{h_j}(x,  \xi_j -1, \xi_{j+1}) &
-F_{h_j} (x, 
\xi_j -1, \xi_{j+1} -1) \\ F_{h_j}(x,  \xi_j ,
\xi_{j+1}) & - F_{h_j}(x, 
\xi_j ,
\xi_{j+1}-1 )
\end{array}\right) \ ,
$$
where 
$$
F_h (x,  \xi, \xi' ) :=
\xi'aY - xbX +  \xi'\xi xc Z -\xi d \ . 
$$
Accordingly, for $\vec{h} \in (\PZ^3)^L$ and  
$\vec{\xi}= (\xi_0,  \ldots , \xi_{L-1} ) \in
(\CZ^N)^L$, the modified
$L$-operator becomes
\begin{eqnarray*}
\widetilde{L}_{\vec{h}}(x,
\vec{\xi} ) : = 
\bigotimes_{j=0}^{L-1}
\widetilde{L}_{h_j}(x,
\xi_j, \xi_{j+1})   = 
\left( \begin{array}{cc}
   \widetilde{L}_{ \vec{h}; 11}(x, \vec{\xi}) &
\widetilde{L}_{
\vec{h};12}(x, \vec{\xi})  \\
     \widetilde{L}_{ \vec{h}; 21}(x, \vec{\xi}) &
\widetilde{L}_{ \vec{h} ; 22}(x, \vec{\xi}) 
\end{array} \right) , \ \ \ \xi_L := \xi_0  \ . 
\end{eqnarray*}
As the procedure of gauge transform keeps 
the same trace, we have 
$ T_{\vec{h}} (x) = {\rm tr}_{aux}
(\widetilde{L}_{\vec{h}}(x,
\vec{\xi} ) ) $. For a given $\vec{h}$, we shall  
consider the variable $(x, \vec{\xi}) $ 
only lying on the curve
${\cal C}_{\vec{h}}$ defined by the system of
equations, 
\begin{eqnarray}
{\cal C}_{\vec{h}} : \ \ \xi_j^N  =(-1)^N
\frac{\xi_{j+1}^Na_j^N - x^Nb_j^N}{\xi_{j+1}^N x^N
c_j^N - d_j^N } \ \ , \ \ \ \  j =0,
\ldots, L-1 . \label{eq:Cvh}
\end{eqnarray} 
The Baxter vacuum
state\footnote{The Baxter vacuum state here is
phrased by the Baxter vector in \cite{FK, LR99}}
over 
${\cal
C}_{\vec{h}}$ is the family of vectors 
$|p
\rangle \in \ 
\stackrel{L}{\otimes} \CZ^N $  with the form 
$$
|p\rangle \ : = |p_0\rangle \otimes \ldots \otimes
|p_{L-1}\rangle 
\in \ \ \stackrel{L}{\otimes} \CZ^N \ , \ \ \ \ p
\in {\cal C}_{\vec{h}} \ ,
$$ 
where $| p_j \rangle$ is the vector in $\CZ^N$
governed by the relation,
\be
\langle 0| p_j \rangle = 1 \ , \ \ \ \
\frac{\langle m|p_j \rangle}{\langle
m-1|p_j \rangle} = 
\frac{\xi_{j+1} a_j
\omega^m  - xb_j }{
- \xi_j (  \xi_{j+1} x c_j \omega^m - d_j ) }  \ .
\ele(Bv)
The constraint of $(x, \vec{\xi})$ on the curve
${\cal C}_{\vec{h}}$ ensures that the following  
properties  hold for the Baxter vacuum state,
\begin{eqnarray*}
\widetilde{L}_{\vec{h}; 1 1}(x,
\vec{\xi})|p\rangle =  |\tau_- p\rangle 
\Delta_-(p) , &
\widetilde{L}_{\vec{h}; 2 2}(x,
\vec{\xi})|p\rangle =  |\tau_+
p\rangle\Delta_+(p) \ , &
\widetilde{L}_{\vec{h}; 2 1}(x,
\vec{\xi})|p\rangle = 0
\ , 
\end{eqnarray*}
where $\Delta_\pm, \tau_\pm$ are (rational)
functions and  automorphisms of 
${\cal
C}_{\vec{h}}$ defined by
\begin{eqnarray}
&\Delta_-(x, \xi_0, \ldots, \xi_{L-1})  &=
\prod_{j=0}^{L-1}( d_j-x
\xi_{j+1} c_j ) \ , \nonumber \\
&\Delta_+(x, \xi_0, \ldots, \xi_{L-1}) &= 
\prod_{j=0}^{L-1} \frac{\xi_j
(a_jd_j-x^2b_jc_j)}{\xi_{j+1}a_j -xb_j} \  , 
\label{DelTau} \\
&\tau_\pm :  (x, \xi_0, \ldots, \xi_{L-1}) 
&\mapsto (q^{\pm 1} x, 
q^{-1} \xi_0, \ldots, q^{-1} \xi_{L-1}) \ 
\nonumber .
\end{eqnarray}
This implies that under the action of 
the transfer matrix, the Baxter vacuum state is
decoupled as the sum of those after the
$\tau_\pm$-transformations:
\bea(l)
T_{\vec{h}}(x) |p \rangle = |\tau_- p\rangle 
\Delta_-(p)  + |\tau_+ p\rangle\Delta_+(p) \ , \ \ 
{\rm for } \ \ p \in {\cal C}_{\vec{h}} \ .
\elea(T|p)
For a common eigenvector $\langle \varphi|
\in \stackrel{L}{\otimes}\CZ^{N*}$ of 
transfer matrices
$T_{\vec{h}}(x)$, the eigenvalue $\Lambda(x) $ is
a polynomial of $x$, i.e., 
$\Lambda(x) \in \CZ[x]$. The function $Q(p)$ on
${\cal C}_{\vec{h}}$ defined by  
$Q(p):= \langle \varphi|p\rangle$ for  $p \in
{\cal C}_{\vec{h}}$  satisfies the
following  Baxter's
$T$-$Q$ relation\footnote{The Baxter $T$-$Q$
relation here was called the  Bethe equation in
\cite{ LR99}}  for
$T_{\vec{h}}(x)$,
\bea(l)
\Lambda(x) Q(p)  = \Delta_-(p) Q(\tau_-(p))  
+ \Delta_+(p) Q(\tau_+(p))  \ , \ \ {\rm for} \ p 
\in {\cal C}_{\vec{h}} \ .
\elea(Bethe)
The above equation is equivalent to the Baxter's
$T$-$Q$ relation for $T^\ast(x)$ with
certain scaling modification of $\Delta_\pm$,
(see \cite{LR01}). The spectral curve
${\cal C}_{\vec{h}}$ is a branched cover the
rational curve,
$x$-complex line. In 
(\req(Bethe)), $\Lambda(x)$ is a polynomial of
$x$, and $Q(p)$ is a rational function on
the Riemann surface ${\cal C}_{\vec{h}}$. For
the solutions of the Baxter's $T$-$Q$ relation
(\req(Bethe)) in  the spectrum problem of the
transfer matrix
$T_{\vec{h}}(x)$, it requires the understanding of
Baxter vacuum state so that one can extract the
necessary properties of zeros and poles of
$Q(p)$, which appears to be a challenging problem
in algebraic geometry    

In this article, we shall concern only the
case $L=3, 4$ for the reason of their special
feature connecting with certain Hamiltonian spin
chains of physical interest. In these cases, the
crux of the transfer matrix
$T(x)$ is the term $T_2$, which can
be converted to the following Hamiltonian. We
won't give the exact identification here, the
details can be found in \cite{LR99, LR01}.

(I) $L=3$. The term $T_2$ is equivalent to the
following Hofstadter type Hamiltonian proposed
by Faddeev and Kashaev \cite{FK}, ( for the
identification using  the convention in this
report, see
\cite{LR99}):
\be
H_{FK}= \mu ( \alpha U + \alpha^{-1} U^{-1} ) + \nu (
\beta V + \beta^{-1} V^{-1}) + \rho ( \gamma W +
\gamma^{-1} W^{-1} ) 
\ele(Hfk)
where $U, V, W$ are unitary operators satisfying the
Weyl commutation relations and the $N$-th power
identity property:
$$
UV=
\omega VU, \ \ VW=
\omega WV, \ \ WU= \omega UW; \ \ U^N= V^N = W^N
=1 \ . 
$$ 
For a particular case of $H_{FK}$ with $\rho=0$,
the system becomes  the Hofstadter
Hamiltonian
\cite{A, H, WZ},
\be
H_{Hof}  =  \mu (   \alpha U +
\alpha^{-1}
 U^{-1})  + \nu ( \beta V +
 \beta^{-1} V^{-1} )  ,
\ele(Hof)
a renowned Bloch electron system with a constant
external magnetic field ( and the
rational magnetic flux in this situation). 

For $L=4$, the interesting physical systems arise
from $T_2^*$ with the following constrains on 
parameters and their relations with the
$C, D$ in $T_0, T_4$,
\bea(ll)
a_jd_j=q^{-1} ,& d_0d_1d_2d_3= 1 ; \\
 b_jc_j=\left\{
\begin{array}{ll}
-k^{-1}
\ ,  &  j=0, 2 \ , 
\\
-k &  j=1, 3 \ ,
\end{array} \right. &   c_1^Nc_3^N = k^{2N}
c_0^Nc_2^N \ ; \\
C = \frac{c_1c_3}{k^2c_0c_2} .
\elea(DQPSG)
With the operators $U_j$,
$$
U_1 = 
Z \otimes X \otimes 1 \otimes 1 , \ \
 U_2 = 1 \otimes Z \otimes X \otimes 1 , \ \   
U_3= 1 \otimes 1 \otimes Z \otimes X , \ \ U_4 = 
X \otimes 1 \otimes 1 \otimes Z ,
$$
one can discuss the following two systems,
(for details, see \cite{LR01}). 

(II) Discrete quantum pendulum. This is the
case (\req(DQPSG)) together with the 
further constraints,
$$
D=1 \ ;  \ \ \ 
 \frac{ d_0 c_1 d_3}{kc_2 } U_2 = 
\frac{k c_0 }{ d_1d_2c_3 }   U_4^{-1} ( = :
Q_{n-1}) \ , \ \ \frac{ c_1}{  kc_0 d_2d_3 }  
U_1^{-1}  = 
\frac{k d_0d_1 c_2 }{c_3} U_3 (=:Q_n )  .
$$
The coefficients of $T^\ast(x)$ in (\req(TT*))
now have the form,
$$
\begin{array}{ll}
\ \ T_0^\ast &= T_4^\ast = 2 \\ [1mm]
-T_2^\ast &= 
2(Q_n +Q_n^{-1} + Q_{n-1}
  + 
Q_{n-1}^{-1}) \\ &
+  k ( qQ_{n-1}Q_n + q^{-1} Q_n^{-1}Q_{n-1}^{-1}) + 
 k^{-1}(   qQ_nQ_{n-1}^{-1}+ q^{-1} Q_{n-1}Q_n^{-1}
) \ .
\end{array}
$$
The above $-T_2^\ast$ is the
Hamiltonian of discrete quantum pendulum
in \cite{BKP}, which is subject to the 
evolution equation, 
$$
Q_{n+1}Q_{n-1} = (\frac{k+qQ_n}{1+qkQ_n})^2 \ , \
\  Q_nQ_{n-1}= q^2 Q_{n-1}Q_n \ .
$$

(III) Discrete sine-Gordon (SG) Hamiltonian.
This is the situation under the condition
(\req(DQPSG)) with one further
identification,
$$
\frac{c_1d_0d_3}{kc_2} D^{\frac{-1}{2}} U_2 =
\frac{kc_0}{c_3d_1d_2} D^{\frac{1}{2}} U_4^{-1} \
\ .
$$
We have 
$$
\begin{array}{ll}
T^{\ast}_0 = & 
D^{\frac{1}{2}} +  D^{\frac{-1}{2}}
\ , \ \ \ \ \ \ \ \ \ \ \ 
T^{\ast}_4  =  \  
D^{\frac{-1}{2}} +  D^{\frac{1}{2}} \ ,
\\ [1mm]
-T^{\ast}_2 = &
\frac{  kc_0 d_2d_3 }{ c_1}   D^{\frac{-1}{2}}U_1
+
\frac{ c_1d_0  d_3}{kc_2 } D^{\frac{-1}{2}}U_2 +
\frac{k  c_2 d_0d_1}{c_3} D^{\frac{-1}{2}}U_3 
  + 
\frac{ c_3d_1d_2 }{k c_0 }D^{\frac{-1}{2}} U_4 
+  \frac{ c_1}{  kc_0 d_2d_3 }  D^{\frac{1}{2}} U_1^{-1} 
\\& + \frac{kc_2 }{  c_1 d_0 d_3} 
D^{\frac{1}{2}} U_2^{-1}  + \frac{c_3}{k
c_2 d_0d_1 } D^{\frac{1}{2}} U_3^{-1} 
 + \frac{k c_0 }{ c_3 d_1d_2 } D^{\frac{1}{2}}
U_4^{-1} 
+  \frac{k
c_1d_0}{qc_3d_2} D^{\frac{-1}{2}}V_1 + \frac{
k qc_3d_2}{c_1d_0}  
D^{\frac{1}{2}} V_1^{-1} \\& + 
\frac{c_0 d_3
}{kqc_2d_1}D^{\frac{-1}{2}} V_4 +\frac{q   
c_2d_1}{kc_0d_3} D^{\frac{1}{2}} V_4^{-1}   \ .
\end{array}
$$
The above $-T_2^*$ can be identified with
the  discrete quantum sine-Gordon integral in
\cite{BKP}.

\section{The  T-Q Polynomial Equation for Rational
Spectral Curves}
In this section we consider the $T$-$Q$ relation
in the  case  when the spectral curve ${\cal
C}_{\vec{h}}$ is totally degenerated into
rational  curves. In this situation, though
geometric features of the spectral data become
trivial, the quest for solutions of the
Baxter's
$T$-$Q$ relation has still raised certain subtle
algebraic problems. We shall first
summarize some basic facts of reducing the
Baxter's $T$-$Q$ relation into a polynomial
equation for an arbitrary size $L$, then feed
the scheme into the case $L=3, 4$ and state 
our results on the models  (I) (II) (III)
of Sect. 2, ( for the detailed account of the
derivation, see \cite{LR99, LR01}).

By the rational degenerated spectral
curve, we mean the  coordinates
$\xi_j^N$ in the equation (\ref{eq:Cvh})
of ${\cal C}_{\vec{h}}$ are constants, independent
of the variable
$x$.  Now the parameter
$h_j$s and the variables $\xi_j$s  are subject to
the constraints:
\be
\frac{b_j^Nd_j^N}{a_j^Nc_j^N} = \frac{a_{j+1}^Nb_{j+1}^N}{c_
{j+1}^Nd_{j+1}^N} \ , \ \ \ \
\xi_j^{2N} = \frac{a_j^Nb_j^N}{c_j^Nd_j^N} \ \ \
\  {\rm for} \ \ 0 \leq j \leq L-1 \ .
\ele(deg)
One special case is given by the relations, $
a_j = q^{-1} d_j $, $b_j = q^{-1} c_j $, 
 and $\xi_j^N=1$ for $ j=0, \ldots, L-1 $
\cite{FK, LR99}. 
For the condition (\req(deg)), we define
$r_j  =
\sqrt{\frac{b_{j-1}d_{j-1}}{a_{j-1}c_{j-1}}} ,
$ for $ j
\in \ZZ_L $.
Then ${\cal C}_{\vec{h}}$ contains the
$\tau_\pm$-invariant curve ${\cal C}$, which
is sufficient for our purpose to studythe Baxter's
$T$-$Q$ equation,
$$
{\cal C} := \{ (x, \xi_0, \ldots, \xi_{L-1}) \ | \ 
 r_0^{-1}\xi_0 = \ldots = 
r_{L-1}^{-1}\xi_{L-1} =  q^l \ , \ \ l \in \ZZ_N
\} \ . 
$$ 
We shall identify ${\cal C}$  with $\PZ^1
\times \ZZ_N$ via the correspondence:
$$
{\cal C} = \PZ^1 \times \ZZ_N \ , \ \ \ \ 
(x, \  r_0 q^l, \ldots, 
r_{L-1} q^l) \longleftrightarrow 
(x, l) \ .
$$
The automorphisms $\tau_\pm$ on ${\cal C}$  become
$\tau_\pm (x, l) = (q^{\pm 1} x, l-1)$, and the
action (\req(T|p)) of $T(x) (:= T_{\vec{h}})$ on
$|x, l\rangle$ now takes the form,
\bea(l)
T (x) |x, l\rangle = |q^{-1}x, l-1\rangle 
\Delta_-(x, l)  + |qx, l-1\rangle\Delta_+(x, l) \  , 
\elea(T|xm)
where $\Delta_\pm$ are 
functions of $x$: 
$$
\frac{\Delta_-(x, l)}{ d_0 \cdots d_{L-1}}= 
\prod_{j=0}^{L-1}( 1-x q^l d_j^{-1}c_j r_{j+1}
),\ \ \ \ 
\frac{\Delta_+(x, l)}{d_0 \cdots d_{L-1}}=
\prod_{j=0}^{L-1} \frac{
1-x^2a_j^{-1}d_j^{-1}b_jc_j}{1 -x q^{-l} a_j^{-1}b_j
r_{j+1}^{-1}}  .
$$
With the substitutions,
$$
(d_0\cdots d_{L-1})^{-1} T(x) \mapsto T(x) \ , \ \ 
(d_0\cdots d_{L-1})^{-1}\Delta_\pm(x, l) \mapsto \Delta_\pm (x, l ) 
\ ,
$$
the relation (\req(T|xm)) still holds for the
modified
$\Delta_\pm$, now with the expression,
$$
\Delta_-(x, l)  = 
\prod_{j=0}^{L-1}( 1-x
   c^*_j q^l) \ , \ \ \ \
\Delta_+(x, l) = 
\prod_{j=0}^{L-1} \frac{
1-x^2c^{* 2}_j}{1 -x c^*_j q^{-l} } \  
$$
where $c_j^*= d_j^{-1}c_j r_{j+1} 
(= a_j^{-1}b_j r_{j+1}^{-1})$. 
Furthermore,  one can convert the expression
(\req(Bv)) of the Baxter vacuum state over the
elements of ${\cal C}$ to the following
component-expression of the Baxter's vector
$|x, l
\rangle$:
\begin{eqnarray*}
\langle {\bf k}|x, l \rangle  = q^{|{\bf k}|^2}
\prod_{j=0}^{L-1} 
\frac{ (x c^*_jq^{-l-2}; \omega^{-1})_{k_j} }{ (x 
c^*_j q^{l+2}; \omega )_{k_j}}
 \ . 
\end{eqnarray*}
Here the bold letter ${\bf k}$ denotes a
multi-index vector
${\bf k}= (k_0, \ldots, k_{L-1})$ for 
$k_j \in \ZZ_N$ with the square-length of ${\bf
k}$ defined by 
$|{\bf k}|^2:= \sum_{j=0}^{L-1} k_j^2$.  
Each ratio-term  in the above right
hand side is given by a non-negative 
representative for each element in $\ZZ_N$ 
appeared in the formula. 
One would like to deduce the
 relation (\req(T|xm)) of the transfer
matrix on ${\cal C}$ to the base 
$x$-line, accordingly conduct the analysis
of the Baxter vacuum state to extract the
essential ingredients for the solutions of the
corresponding Baxter's $T$-$Q$ relation. Indeed,
by the Fourier transform method  on the Baxter
vacuum state of ${\cal C}$, then applying certain
"diagoalization" process of the transfer matrix
on the modified Baxter vacuum state, one obtains
the following results, (for  the detailed
derivation , see
\cite{LR99, LR01}).
\begin{theorem}  \label{thm:LR}
 Let $f^e, f^o$ be the functions on 
${\cal C}$,
\begin{eqnarray*}
f^e(x, 2n) :=  \prod_{j=0}^{L-1}
\frac{(xc^*_j ; \omega^{-1} )_{n+1} }{
(xc^*_j; \omega )_{n+1}} , 
\ \ \
f^o(x, 2n+1) :=  \prod_{j=0}^{L-1}
\frac{(xc^*_jq^{-1}; \omega^{-1} )_{n+1}}{
(xc^*_jq ; \omega)_{n+1}} \ .
\end{eqnarray*}
For $x \in \PZ^1,  l \in \ZZ_N$,
we define the following vectors in
$\stackrel{L}{\otimes}\CZ^N$, 
\begin{eqnarray*}
|x\rangle_l^e = \sum_{n=0}^{N-1}  |x, 2n\rangle  
f^e (x, 2n)
\omega^{ln} 
 \ , &
|x\rangle_l^o = \sum_{n=0}^{N-1} 
 |x, 2n+1\rangle f^o (x, 2n+1)  \omega^{ln}  \ , 
\\
|x \rangle_l^+ =  |x\rangle_l^e q^{-l} u(qx) +  
|x \rangle_l^o u(x) & {\rm where} \ \ 
u(x): = \prod_{j=0}^{L-1} (1-x^Nc^{*
N}_j)(xc^*_jq ; q^2)_M \ .
\end{eqnarray*} 
Then 

(i) $|x \rangle_l^e u(qx) = 
|x\rangle_l^o q^lu(x)$, or equivalently, 
$
|x\rangle_l^+ = 2 q^{-l} |x\rangle_l^e u(qx)  = 2
|x\rangle_l^ou(x) $. \\[.1mm]

(ii) The $T(x)$-transform on $| x\rangle^+_l$, $ l \in
\ZZ_N$,
is given by
$$
q^{-l} T(x)|x\rangle^+_l
=|q^{-1}x\rangle^+_l  \Delta_-(x, -1) +
|qx\rangle^+_l  \Delta_+(x, 0)  \ .
$$

(iii) For a common eigenvector $\langle \varphi|$
of
$T(x)$ with the  eigenvalue
$\Lambda (x)$, the function
$Q_l^+(x)  (:=\langle \varphi | x\rangle^+_l)$
and $\Lambda (x)$ are  polynomials with the
properties:
$$
\begin{array}{llll}
{\rm deg.}Q^+_l (x)
\leq  (3 M +1)L , &
 {\rm deg.}
\Lambda (x) \leq 2[\frac{L}{2}], &
\Lambda (x)= \Lambda(-x), & 
\Lambda (0)= 
q^{2l}+1  ,
\end{array}
$$
and the following Baxter's $T$-$Q$ equation
holds:
\begin{eqnarray*}
 q^{-l} \Lambda (x) Q_l^+ (x) = 
\prod_{j=0}^{L-1}(1-xc^*_jq^{-1})
Q_l^+ ( xq^{-1}) + \prod_{j=0}^{L-1} 
(1+x c^*_j) Q_l^+ (xq) \ .
\end{eqnarray*}
Furthermore for $0 \leq m \leq M$, 
$Q_m^+(x), Q_{N-m}^+(x) \in  x^m \prod_{j=0}^{L-1}
(1-x^Nc^{* N}_j)
\CZ[x]$. 
\end{theorem}
$\Box$ \par \vspace{0.2in} \noindent
By $(iii)$ of the above theorem,
the  Baxter's $T$-$Q$ equations  for
the sectors
$m, N-m$ can be unified into a single one. 
From now on, we shall always use the letter $m$
to denote an integer between 
$0$ and $M$,
$$
0 \leq m \leq M \ . 
$$
By introducing the 
polynomials
$\Lambda_m(x), Q ( x)$ via the relation,
$$
(\Lambda_m(x) , \ x^m
\prod_{j=0}^{L-1} (1-x^Nc^{* N}_j) Q ( x) ) = 
(q^{-m}\Lambda(x), \ Q_m^+(x) ) , \ \ (q^m
\Lambda (x) , \ Q_{N-m}^+(x) )  \ ,
$$
the equations in Theorem \ref{thm:LR} $(iii)$ for
$l=m, N-m$,  are equivalent to the  following 
polynomial equation of $Q (x), \Lambda_m(x)$: 
\be
 \Lambda_m (x) 
Q (x) = q^{-m} 
\prod_{j=0}^{L-1}(1-xc^*_jq^{-1}) 
Q (xq^{-1})+ q^m \prod_{j=0}^{L-1} (1+x c^*_j) 
 Q (xq) \ ,
\ele(rBeq)
with the following constraints of $Q(x)$ and $
\Lambda_m(x)$,
$$
{\rm deg.}Q (x)
\leq ML-m ,  \ \ 
 {\rm deg.}
\Lambda_m (x) \leq 2[\frac{L}{2}], \ \ 
\Lambda_m (x)= \Lambda_m (-x),  \ \ \Lambda_m
(0)=  q^m + q^{-m}. 
$$
By (\req(TT*)), the above $\Lambda_m(x)$ is
indeed the eigenvalue of $T^{\ast}(x) $, and 
(\req(rBeq)) corresponds the Baxter's $T$-$Q$
equation for
$T^{\ast}(x)$ in the sectors
$m, N-m$.

We now consider the solutions of the equation
(\req(rBeq)) for
$L=3, 4$ for the models (I) (II) (III)
in Sect. 2. We shall consider the 
$c_j^*$s to be generic. Furthermore, for $L$=4, 
the parameter
$\vec{h}$ will be  confined in the 
regime,
$$
qa_j= d_j= 1 \ , \ \ \ 
-b_j=c_j= \left\{\begin{array}{ll}
k^{\frac{-1}{2}}& {\rm for \ even} \ j , \\
k^{\frac{1}{2}} &{\rm for \ odd} \ j ,
\end{array}
\right. 
$$
and the eigenvalue
$\Lambda_m(x)$ in (\req(rBeq)) will be a
reciprocal polynomial, 
$$
\Lambda_m (x)= q^m+q^{-m} + \lambda x^2 +
(q^m+q^{-m}) x^4
\ ,
$$
i.e., $\Lambda_m(x) = \Lambda_m^\dagger(x)$. Here
for $P(x) \in \CZ[x] $,
$P^\dagger(x)$ is the polynomial defined by
$$
P^\dagger(x) = x^{{\rm deg}(P)} P(x^{-1}) \ .
$$
 The  Baxter's
$T$-$Q$ polynomial equation with 
a reciprocal $\Lambda_m(x)$ will be called a
symmetric one. Note that for the discrete quantum
pendulum (II) and discrete SG Hamiltoinian (III)
in the above regime, we have
$C=1$ by  (\req(DQPSG)), hence the
reciprocal property of
$\Lambda_m(x)$ always hold.  The
following results were shown in 
Theorem 3 of \cite{LR99} and Theorem 3
\cite{LR01}:
\begin{theorem} \label{thm:RTQ34}
(i) For $L=3$, the Baxter's
$T$-$Q$ polynomial equation
$(\req(rBeq))$ has
$N$ distinct eigenvalues
$\Lambda_m(x)$, each of which has 
one-dimensional eigenspace generated by a monic 
eigen-polynomial
$Q(x)$ of degree $d=3M-m$ with $Q(0) \neq 0$. In
this case, the corresponding Hofstadter type
Hamiltonian $(\req(Hfk))$ has the
parameters
$\alpha=\beta= \gamma =1$.
 
(ii) For $L=4$ and the symmetric Baxter's
$T$-$Q$ polynomial equation
$(\req(rBeq))$,  there
are
$N$ distinct eigenvalues
$\Lambda_m(x)$, each of which  has 
one-dimensional eigenspace generated by a monic 
eigen-polynomial
$Q(x)$ of degree $d=4M-2m$ with $Q(0) \neq 0$ and
$Q^\dagger(x) = \pm Q(x)$. Furthermore among
these $N$ eigen-polynomials, there are  
$M+1$ of $Q(x)$s with the type
$Q^\dagger(x) = Q(x)$, and the other $M$ ones
are of the type $Q^\dagger(x) = - Q(x)$. In
particular, the Baxter's $T$-$Q$
polynomial relation of the SG
model is described by the sector $C=1$ for all
sectors
$m$, and the discrete quantum pendulum is the one
with $C=1$, 
$m=0$.
\end{theorem}
$\Box$ \par \vspace{0.2in} \noindent
Now we discuss the relation between the $T$-$Q$
polynomial  equation (\req(rBeq)) and the usual 
Bethe Ansatz formulation in the physical
literature. In (\req(rBeq)) when $Q(0) \neq 0$,
which is the case in Theorem \ref{thm:RTQ34}, one
can write 
$$
Q (x) = \prod_{i=1}^d (
x -
\frac{1}{z_i} ) \ , \ \ \ 
\  z_i \in \CZ^*  \ .
$$
Substituting $z_l^{-1}$ in (\req(rBeq)), the 
$z_i$s satisfy the following Bethe
Ansatz relations,
\be
 q^{L+2m+d} \prod_{j=0}^{L-1}
\frac{z_l+ c^*_j } {qz_l-
c^*_j }
 = \prod_{i=1, i \neq k }^d \frac{  z_i - q z_l
 }{ q  z_i- z_l } \ , \ \ l=1, \ldots,  d \ .
\ele(BT)
For the Hofstadter type Hamiltonian (\req(Hfk)),
by Theorem \ref{thm:RTQ34} $(i)$ the Bethe Ansatz
(\req(BT)) for the sector $m$  has the form 
$$
 q^{m+\frac{3}{2}}  \prod_{j=0}^2 \frac{z_l +
 c_j^*}{qz_l - c_j^*} =   \prod_{i=1, i \neq
k}^{3M-m}
\frac{ z_i -  qz_l }{ 
 q z_i -z_l  } \ ,  \ \ 1 \leq l \leq 3M-m \ .
$$
The above relation for the sector $m=M$ is the
one  postulated in \cite{FK}, where in
some special case, one can  reproduce
the spectrum of Hofstadter Hamiltonian found in
\cite{WZ}. For the $SG$ model and the discrete
quantum pendulum with the parameter $k$, by
Theorem
\ref{thm:RTQ34}
$(ii)$, the Bethe Ansatz (\req(BT)) now takes the
form 
\begin{eqnarray*}
 (\frac{z_l^2 + 2 {\rm i}c
q^{\frac{1}{2}} z_l -  q}{ q z_l^2 - 2  {\rm i} c
q^{\frac{1}{2}}z_l
 - 1})^2 = \prod_{ i=1, i \neq l}^{4M-2m} 
\frac{z_i -q z_l}{qz_i - z_l} \ , \ \ \ \ \ 1 \leq
l \leq 4M-2m \ ,
\end{eqnarray*}
where $c =
\frac{1}{2}(k^{\frac{1}{2}}+k^{\frac{-1}{2}})$.
The solutions of the above Bethe Ansatz represent
the  spectra for those model arising from
the rational degenerated spectral curve.

\section{The General Spectral Curve for
Hofstadter Model,  Discrete Quantum Pendulum and
Discrete Sine-Gordon Model} 
We are going to discuss certain geometrical
aspect of the Hofstadter, 
discrete quantum pendulum and SG models in a
general spectral curve case. Now the curves ${\cal
C}_{\vec{h}}$ (\ref{eq:Cvh}) are Riemann
surfaces with a very high genus. However, the
values of
$\xi_j^N$s for 
${\cal C}_{\vec{h}}$ are determined only by 
the variables 
$\eta (:=\xi_0^N), y (:= x^N)$, which defines the
curve 
$$
{\cal B}_{\vec{h}} : \ \ \ C_{\vec{h}}(y) \eta^2 +
(A_{\vec{h}}(y)-  D_{\vec{h}}(y))\eta 
- B_{\vec{h}}(y) = 0 \ ,
$$
where $A_{\vec{h}}, B_{\vec{h}}, 
C_{\vec{h}}, D_{\vec{h}}$ are polynomials of
$y$ expressed by the relation, 
$$
\left( \begin{array}{cc}
-A_{\vec{h}}(y)  &  B_{\vec{h}}(y)  \\
C_{\vec{h}}(y)  & - D_{\vec{h}}(y)    
\end{array} \right) :
= \prod_{j=0}^{L-1}  \left( \begin{array}{cc}
- a_j^N  &  y b_j^N \\
 y c^N_j  &-d^N_j    
\end{array} \right) \ .
$$
The curve ${\cal C}_{\vec{h}}$ becomes a
(branched) cover of ${\cal B}_{\vec{h}}$.

For the Hofstadter model (\req(Hof)), it is
the case when $L = 3 $ with $ a_0=d_0=0 \ , \
b_0=c_0 = 1 $ and  $h_1, h_2$ generic. For the 
discrete quantum pendulum and $SG$ model, the
constraints of the parameter $\vec{h}$ are given
those given in (\req(DQPSG)). In all these cases,
the curves ${\cal B}_{\vec{h}}$ form a family of
elliptic curves, covered by the spectral
family ${\cal C}_{\vec{h}}$. The
geometrical feature of these covering families
has been explored in cooperation with the
Baxter's $T$-$Q$ relation on ${\cal C}_{\vec{h}}$,
but only on the qualitative aspect \cite{LR99,
LR01}. Nevertheless, the
quantitative study of the
$T$-$Q$ relation, which is the core of the
problem, has still remained a difficult task,due
to the lack of proper understanding of the Baxter
vacuum state over
${\cal C}_{\vec{h}}$. The prospect of employing
the elliptic function theory reated to the
family ${\cal B}_{\vec{h}}$ to the
Baxter $T$-$Q$ relation would
be a challenging problem, which we leave to future
work.

\section{Further Remarks}
We have made a short review of our recent results
on  Hofstadter type models, the quantum pendulum
and the discrete sine-Gordon model by the transfer
matrix technique within the frame work of quantum
inverse scattering method. The diagonalization
problem of Hamiltonians for these
model relies on solving the Baxter's $T$-$Q$
relation over a spectral curve. This relation 
arises from the
 transfer matrix (of a finite size
$L$) on the Baxter's vacuum state, to which we
have conducted a systematic study from algebraic
geometry aspect. The complexity of the spectral
curve, consequently the solvability of its related
$T$-$Q$ equation, depends on the length $L$ and
the parameters appeared in the Hamiltonian. For
the three models which we consider in this
report, the transfer matrix has the size $L= 3,
4$, and the spectral curves with generic
parameters are  all related to elliptic
curves. When the spectral curve is totally
degenerated into rational curves, the Baxter's
$T$-$Q$ relation can be converted into a
polynomial equation, in which case we have
obtained a complete solutions of the
diagonalization problem of these models. The
rigorous mathematical derivation enables us to
have a better understanding on both the
qualitative and quantitative nature of the
Baxter's $T$-$Q$ polynomial equation, and Bethe
Ansatz incorporating the relevant physical
consideration. Indeed, our study
in this context strongly suggests that some
further mathematical feature of the  Baxter's
$T$-$Q$ polynomial equation could possibly appear
as a certain 
$q$-Strum-Liouville problem at roots of unity
$q^N=1$. 
Such program is now under investigation, the
results of which we hope to report in near future.

\end{document}